\documentclass[a4paper,11pt]{article}
\usepackage{pos}

\title{Large-$N$ limit as a second quantization}

\author*[a]{Masanori Hanada}

\affiliation[a]{Department of Mathematics, University of Surrey, Guildford, Surrey, GU2 7XH, UK}

\emailAdd{m.hanada@surrey.ac.uk}

\abstract{I propose a simple geometric interpretation for gauge/gravity duality that relates the large-$N$ limit of gauge theory to the second quantization of string theory. }

\FullConference{%
  Corfu Summer Institute 2021 "School and Workshops on Elementary Particle Physics and Gravity"\\
  29 August - 9 October 2021\\
  Corfu, Greece
}

\begin{document}
\maketitle
\section{Introduction}
Gauge/gravity duality \`{a} la Maldacena~\cite{Maldacena:1997re} is a promising framework for the study of non-perturbative aspects of quantum gravity. 
An important open problem is how the bulk geometry of string theory is encoded in gauge theory. 
Given that the duality can work even for the matrix model, there is no doubt that the color degrees of freedom play crucial roles. 

\begin{figure}[htbp]
\begin{center}
\scalebox{0.2}{
\includegraphics{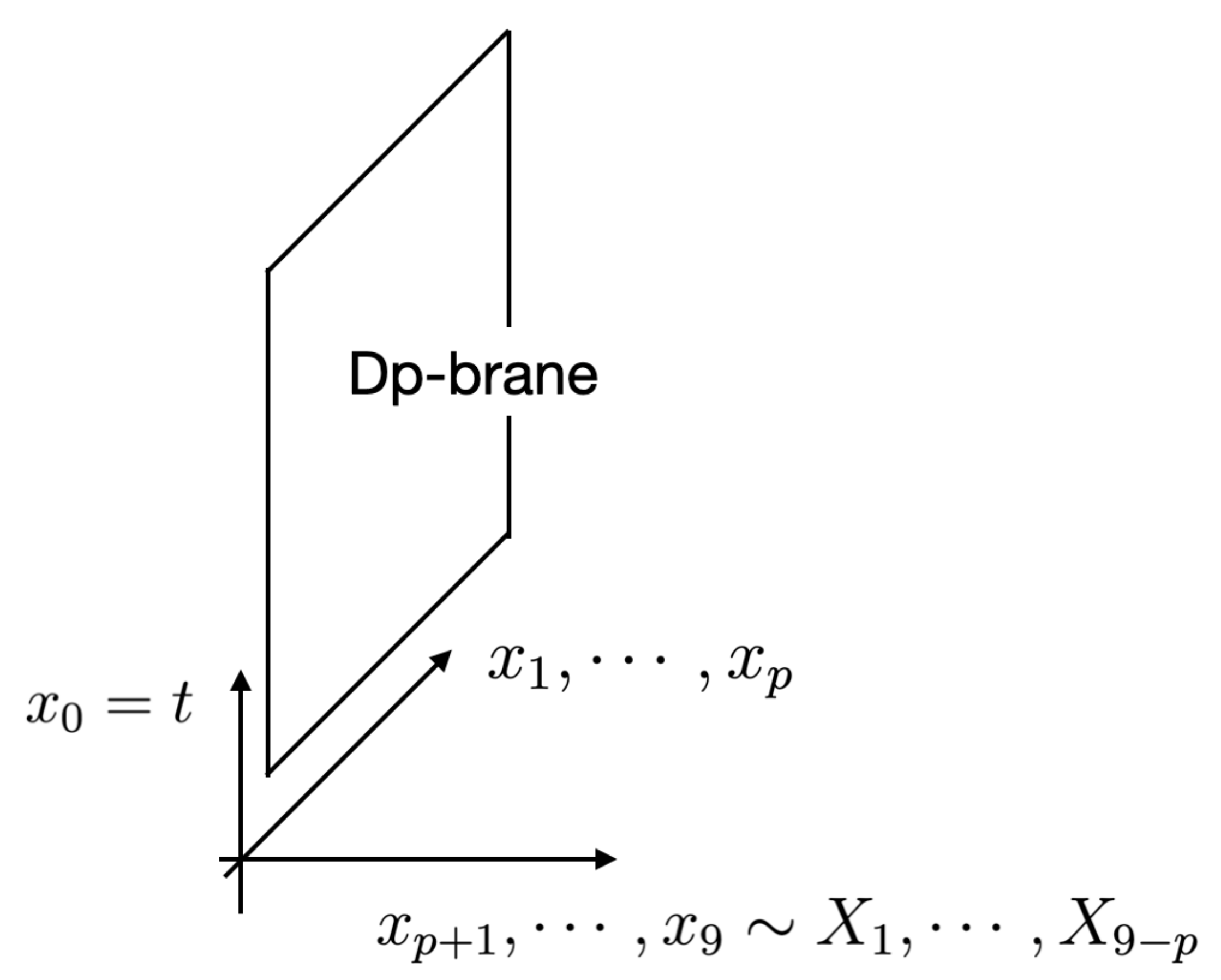}}
\end{center}
\caption{
Emergent space from gauge-theory degrees of freedom, when $(p+1)$-dimensional maximally-supersymmetric SU($N$) Yang-Mills theory is seen as the low-energy effective action of D$p$-branes and open strings~\cite{Witten:1995im}. 
D$p$-branes are extended along $(x_0,x_1,\cdots,x_p)$ in $\mathbb{R}^{1,9}$. The transverse directions $\mathbb{R}^{9-p}$, parametrized by $(x_{p+1},\cdots,x_9)$, are described by adjoint scalars $X_1,\cdots,X_{9-p}$. 
}\label{fig:Dp-brane}
\end{figure}

In a somewhat different context, a natural mechanism of emergent geometry has been known. 
When $(p+1)$-dimensional maximally-supersymmetric SU($N$) Yang-Mills theory is regarded as the low-energy effective theory describing $N$ D$p$-branes and open strings between them, $9-p$ adjoint scalar fields $X_I (I=1,\cdots,9)$ contain the information of transverse $\mathbb{R}^{9-p}$ (Fig.~\ref{fig:Dp-brane})~\cite{Witten:1995im}. The action contains the interaction term $\int d^{p+1}x{\rm Tr}[X_I,X_J]^2$ that favors the simultaneously-diagonalizable configurations.
When all scalars are close to diagonal, i.e., $X_{I,ij}\simeq x_{I,i}\delta_{ij}$, the $(i,i)$-elements $\vec{x}_i=(x_{1,i},\cdots,x_{9-p,i})\in\mathbb{R}^{9-p}$ are regarded as the coordinates of $i$-th D-brane. The off-diagonal elements correspond to the open-string excitations, i.e., if the $(i,j)$-elements are large then a lot of strings are excited between the $i$-th and $j$-th D-branes. 

The Matrix-Theory conjecture~\cite{Banks:1996vh} pushes this interpretation further in the case of $p=0$, and claims this `eigenvalue = location' picture defines the target space in a certain limit of M-theory.
If $M\times M$ sub-matrices take nontrivial values and do not commute with each other, they would describe extended objects such as a black hole.  
In the large-$N$ limit, various many-body states are realized as block-diagonal matrix configurations. 
It suggests a natural possibility of relating the large-$N$ limit of the matrix model and the second quantization of the gravity dual.

There was widespread skepticism regarding the applicability of this interpretation to gauge/gravity duality for a few reasons. Firstly, \\
\begin{center}
(i) There was a convincing argument against the `eigenvalue = location' picture~\cite{Polchinski:1999br,Susskind:1998vk,Heemskerk:2012mn}.  
\end{center}
Furthermore, 
\begin{center}
(ii) No mechanism for the emergence of block-diagonal matrices at strong coupling was known.
\end{center}
Note also that
\begin{center}
(iii) `Nothing' and `something' cannot be distinguished, because the sizes of the blocks have to sum up to $N$. In this sense, although the Matrix-Theory interpretation admits various multi-body states, it is not a genuine second quantization yet.   
\end{center} 
In this talk, I resolve these issues and propose an extremely simple geometric picture for gauge/gravity duality that relates the large-$N$ limit of gauge theory to a second quantization of gravity. 
\section{Geometry from wave packets}
The logic of Refs.~\cite{Polchinski:1999br,Susskind:1998vk,Heemskerk:2012mn} was as follows. 
Consider the action of the form 
\begin{eqnarray}
S
=
\int d^{p+1}x
{\rm Tr}\left(
\frac{1}{2}
F_{\mu\nu}^2
+
\frac{1}{2}
(D_\mu X_I)^2
+
\frac{g^2}{4}
[X_I,X_{J}]^2 
+
{\rm fermion\ part}
\right),  
\nonumber
\end{eqnarray}
where $F_{\mu\nu}$ is the field strength and $D_\mu X_I$ is the covariant derivative. 
We focus on the 't Hooft large-$N$ limit ($\lambda=g^2N$: fixed, the energy $E\sim N^2$).\footnote{Other parameter regions can be studied in similar manners.} 
Via the 't Hooft counting, we obtain $\langle {\rm Tr}X_I^2\rangle\sim N^2$. 
Hence typical eigenvalues of $X_I^2$ and $X_I$ should scale as $N$ and $\sqrt{N}$, respectively.
This scaling is not expected from a dual gravity point of view; for $p<3$, the entire bulk region where the weakly-curved gravity picture is valid is covered, and for $p=3$, the sub-AdS region is covered. 
Furthermore, a similar argument shows that the commutator $[X_I,X_J]$ is far from zero. 
The noncommutativity is so large that, when $X_1$ is diagonalized, $X_{2,3,\cdots}$ are not close to diagonal at all. 
Hence D-branes and open strings form a big fuzz filling a large portion of the bulk geometry. This argument uses only the 't Hooft counting, and hence, it applies to the low-energy states including the ground state.

Ref.~\cite{Hanada:2021ipb} showed this argument is not correct, and resolved the issue (i). 
Let us take the D0-brane theory as an example.
(Generalization to other theories is straightforward.)
The Hamiltonian is given by   
\begin{eqnarray}
\hat{H}
=
{\rm Tr}\left(
\frac{1}{2}
\hat{P}_I^2
-
\frac{g^2}{4}
[\hat{X}_I,\hat{X}_{J}]^2 
+
{\rm fermion\ part}
\right).   
\nonumber
\end{eqnarray}
Each `matrix' $\hat{X}_{I=1,\cdots,9}$ consists of $N^2$ operators, and $\hat{X}$ and conjugate momentum $\hat{P}$ satisfy the canonical commutation relation.
The states mapped to each other via SU($N$)-transformations are identified.\footnote{
This is equivalent to considering the gauge-invariant Hilbert space. See the following discussion and Ref.~\cite{Hanada:2021ipb}. 
}
Refs.~\cite{Polchinski:1999br,Susskind:1998vk,Heemskerk:2012mn} assumed the existence of c-number matrices $X_{I=1,\cdots,9}$ that satisfy ${\rm Tr}X_I^2=\langle{\rm Tr}\hat{X}_I^2\rangle\sim N^2$. 
Implicitly, they assumed a coordinate eigenstate $|X\rangle$ that satisfies $\hat{X}_{I,ij}|X\rangle=X_{I,ij}|X\rangle$ for all $I,i,j$. 
This assumption is not legitimate because low-energy states are not coordinate eigenstates.\footnote{
Typically, such a coordinate eigenstate is considered because of the analogy to the master field at large $N$. It is true that quantities such as $\langle{\rm Tr}\hat{X}_I^2\rangle$ can be precisely reproduced by taking a generic coordinate eigenstate from the wave packet, but such quantities are not useful for identifying the geometry, as Polchinski already speculated when he suggested the puzzle. 
}
Rather, we should consider wave packets in $\mathbb{R}^{9N^2}$. 
The `matrices' that carry the information of D-branes and strings are the center of a wave packet, which we denote by $N\times N$ matrices $Y_I$ ($I=1,2,\cdots,9$). 
The location of the wave packet $Y_I$ can move in $\mathbb{R}^{9N^2}$ via gauge transformation as $Y_I\to UY_IU^{-1} (U\in{\rm SU}(N))$, but the size and shape do not change. The distribution of D-branes determined by the center of the wave packet is well localized. In particular, there is a ground state in which all D-branes are at the origin of $\mathbb{R}^{9}$ and no open strings are excited, that is gauge-invariant and localized around the origin of $\mathbb{R}^{9N^2}$, i.e., $Y_I=0$.  
Various well-localized wave packets around different points in $\mathbb{R}^{9N^2}$ exist, and the overlap between two wave packets remains small if the centers of wave packets are separated more than an $O(N^0)$ value. 

\begin{figure}[htbp]
\begin{center}
\scalebox{0.15}{
\includegraphics{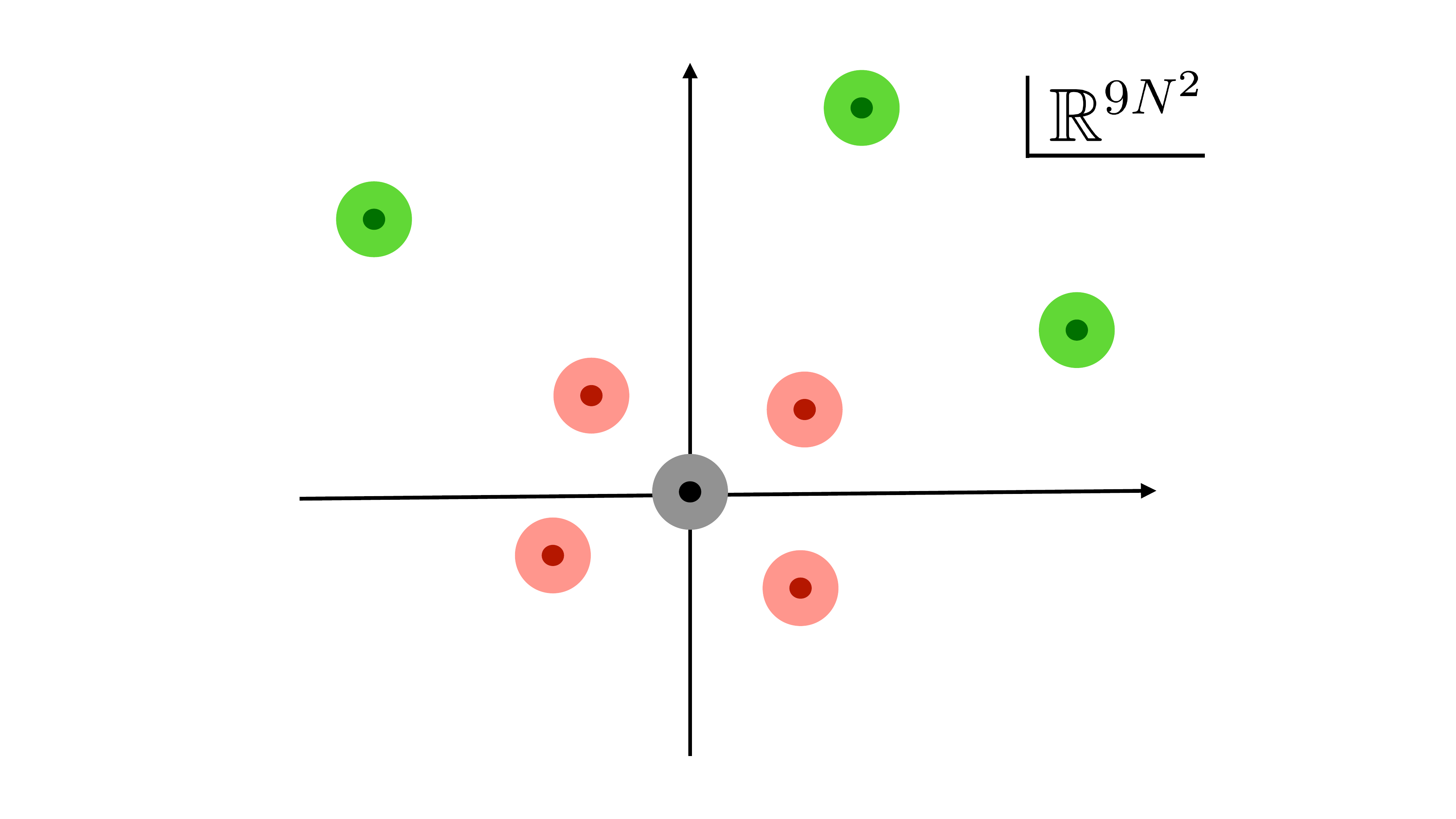}}
\end{center}
\caption{
A visual sketch of wave packets (disks with light colors) in $\mathbb{R}^{9N^2}$. Although the center of a wave packet $Y_I$ (points with thick colors) can move in $\mathbb{R}^{9N^2}$ via the gauge transformation, the shape does not change. 
There are SU($N$)-invariant states, including a single-bunch ground state in which all D-branes are sitting at the origin of $\mathbb{R}^{9}$. 
Wave packets related via the SU($N$) transformation, which are regarded as the same state, are shown with the same color. 
}\label{fig:Gauge-Transformation}
\end{figure}

To understand the properties of wave packets described above, it is instructive to see the simplest and almost trivial case: gauged Gaussian matrix model. 
The Hamiltonian is 
\begin{eqnarray}
\hat{H}_{\rm Gaussian}
=
\sum_{I=1}^9
{\rm Tr}\left(
\frac{1}{2}
\hat{P}_I^2
+
\frac{1}{2}
\hat{X}_I^2
\right)
=
\frac{1}{2}
\sum_{I=1}^9\sum_{\alpha=1}^{N^2}\left(\hat{P}_{I,\alpha}^2+\hat{X}_{I,\alpha}^2\right), 
\nonumber
\end{eqnarray}
where $\alpha=1,2,\cdots,N^2$ is the adjoint index and the canonical commutation relation is given by $[\hat{X}_I^\alpha,\hat{P}_J^\beta]=i\delta_{IJ}\delta^{\alpha\beta}$. 
Namely, we consider $9N^2$ harmonic oscillators subject to the SU($N$) gauge symmetry. 
This model possesses all the problems pointed out in Refs.~\cite{Polchinski:1999br,Susskind:1998vk,Heemskerk:2012mn}; 
following their argument, we would conclude that the low-energy wave functions delocalize. 
We can easily see how this conclusion is falsified, because this model is solvable. 

The ground state is simply the tensor product of $9N^2$ Fock vacua, 
\begin{align}
|0\rangle
\equiv
\otimes_{I,\alpha}|0\rangle_{I,\alpha}.
\end{align}
The wave function in the coordinate representation is 
\begin{align}
\langle X|0\rangle
=
\frac{1}{\pi^{9N^2/4}}
\exp\left(
-\frac{1}{2}\sum_I{\rm Tr}X_I^2
\right). 
\end{align}
This state is SU($N$)-invariant and well-localized about the origin of $\mathbb{R}^{9N^2}$.  
Other wave packets defined above are the coherent states, 
\begin{align}
|{\rm W.P.};Y,Q\rangle
=
e^{-i\sum_I{\rm Tr}(Y_I\hat{P}_I-Q_I\hat{X}_I)}|0\rangle. 
\end{align}
Under the gauge transformation $\hat{X}_I,\hat{P}_I\to U^{-1}\hat{X}_IU,U^{-1}\hat{P}_IU$, this wave packet transforms as  
\begin{align}
|{\rm W.P.};Y,Q\rangle
\to
|{\rm W.P.};Y^{(U)},Q^{(U)}\rangle
\equiv
|{\rm W.P.};UYU^{-1},UQU^{-1}\rangle.
\end{align}
The wave function is well localized around $Y_I$ in the coordinate space and $Q_I$ in the momentum space. For example, when $Q_I$ is zero, 
\begin{align}
\langle X|{\rm W.P.};Y,Q=0\rangle
=
\frac{1}{\pi^{9N^2/4}}
\exp\left(
-\frac{1}{2}\sum_I{\rm Tr}(X_I-Y_I)^2
\right). 
\end{align}
The overlap between two wave packets is
\begin{align}
\langle {\rm W.P.};Y',Q'=0|{\rm W.P.};Y,Q=0\rangle
=
\exp\left(
-\frac{1}{8}\sum_I{\rm Tr}(Y_I-Y'_I)^2
\right). 
\end{align}
If we take $Y_{I,ij}'=0$ (i.e., ground state) and $Y_{I,ij}=y_I\delta_{iN}\delta_{j,N}$ (i.e., $N$-th D-brane is located at $\vec{y}=(y_1,\cdots,y_9)$), the overlap between two wave functions defined on $\mathbb{R}^{9N^2}$ is small as long as $|\vec{y}|\gg 1$. 
Therefore, the resolution of the `location' is of order $N^0$, rather than order $N^{1/2}$. 

Let us consider the case that $\vec{Y}={\rm diag}(\vec{y}_1,\cdots,\vec{y}_N)$, where the vector symbol represent 9-vector. 
The $(i,j)$-component receives a mass proportional to $|\vec{y}_i-\vec{y}_j|$ due to the Higgsing. Furthermore, when some $\vec{y}_i$'s coincide, symmetry of the state is enhanced. Therefore, $\vec{y}_i$ should be regarded as the location of the $i$-th D-brane, following Witten's interpretation~\cite{Witten:1995im}.
Note that we have to identify $Y_I$, and not other points in the wave packet, with the matrices describing the D-brane geometry, for the logic given in Ref.~\cite{Witten:1995im} to work. 
\section{Why block diagonal?}
To resolve the issues (ii) and (iii), a close connection~\cite{Hanada:2020uvt} between color confinement in the large-$N$ gauge theory and Bose-Einstein condensation~\cite{einstein1924quantentheorie} plays a crucial role.   
A trivial but important point is that quantum mechanics of $N$ indistinguishable bosons is a special case of gauge theory, 
whose gauge group S$_N$ is permutations of particles~\cite{Hanada:2020uvt,feynman_superfluidity1}.

The canonical partition function of gauge theory with gauge group $G$ at temperature $T$ is 
\begin{align}
Z(T)=\frac{1}{{\rm vol}(G)}\int_G dg{\rm Tr}(\hat{g}e^{-\hat{H}/T}),
\nonumber
\end{align} 
where the trace is taken over the extended Hilbert space containing non-singlet modes. 
(The micro-canonical partition function can be obtained by constraining the energy.)  
The operator $\hat{g}$ is the gauge transformation corresponding to $g\in G$, 
and $\frac{1}{{\rm vol}(G)}\int_G dg \hat{g}$ is the projector to the singlet sector.\footnote{ 
The integral is taken over the Haar measure. When $G$ is discrete, a sum is taken instead of the integral.}$^,$\footnote{This $\hat{g}$ corresponds to the Polyakov loop in the Euclidean path integral formalism of finite-temperature theory~\cite{Hanada:2020uvt}. Therefore, we can define the Polyakov loop for the system of $N$ indistinguishable bosons, and characterize the Bose-Einstein condensation in terms of Polyakov loop, as for confinement in gauge theory~\cite{Hanada:2020uvt}.}

Let us consider the free limit of the system of $N$ indistinguishable bosons. 
The energy-einegstates are products of one-particle eigenstates, schematically expressed as 
$|\phi_1,\phi_2,\cdots,\phi_N\rangle$. A permutation $g\in G={\rm S}_N$ transforms such a state to $|\phi_{g(1)},\phi_{g(2)},\cdots,\phi_{g(N)}\rangle$. Generic excited states are not invariant under any nontrivial permutation, and hence, only $g=\textbf{1}$ has a nonzero contribution to the partition function. On the other hand, the ground state is invariant under any permutation, 
and there is an enhancement factor $N!$ that corresponds to ${\rm vol}(G)$.
This enhancement triggers the Bose-Einstein condensation~\cite{einstein1924quantentheorie}, and the Bose-Einstein condensate is characterized by the invariance under S$_N$~\cite{feynman_superfluidity1}.   
In the same manner, the confining ground state of ${\rm SU}(N)$ gauge theory is invariant under $G={\rm SU}(N)$, 
and receives the enhancement factor ${\rm vol}(G)$ that triggers the color confinement~\cite{Hanada:2020uvt}.\footnote{
More precisely, $G$ is the set of all ${\rm SU}(N)$-valued functions $g(\vec{x})$, where $\vec{x}$ is spatial coordinates in QFT. 
} 

This mechanism of color confinement, combined with the `eigenvalue = location' interpretation, naturally resolves the issues (ii) and (iii). 
For concreteness, let us consider a large-$N$ confining gauge theory with the first-order confinement/deconfinement transition, 
e.g., 4d Yang-Mills on S$^3$. 
To deconfine all colors, the energy of order $N^2$ is needed. 
What if $E\sim \epsilon N^2$, where $\epsilon$ is of order $N^0$ but much smaller than $1$?
Then the excitations coalesce to form an $M\times M$ block (i.e., $Y_{I,ij}$ is nonzero at $1\le i,j\le M$), where $M\sim \sqrt{\epsilon}N$.
This is because a large symmetry SU($N-M$) can be preserved in this way,
which leads to a large enhancement factor ${\rm vol}({\rm SU}(N-M))$. 
This SU($M$)-partial-deconfinement \cite{Hanada:2016pwv,Berenstein:2018lrm,Hanada:2018zxn}
can occur both at weak-~\cite{Hanada:2019czd} and strong-coupling regions~\cite{Watanabe:2020ufk}. 
In gauge/gravity duality, the partially-deconfined phase is a natural counterpart of the small black hole~\cite{Hanada:2016pwv}.

\begin{figure}[htbp]
\begin{center}
\scalebox{0.5}{
\includegraphics{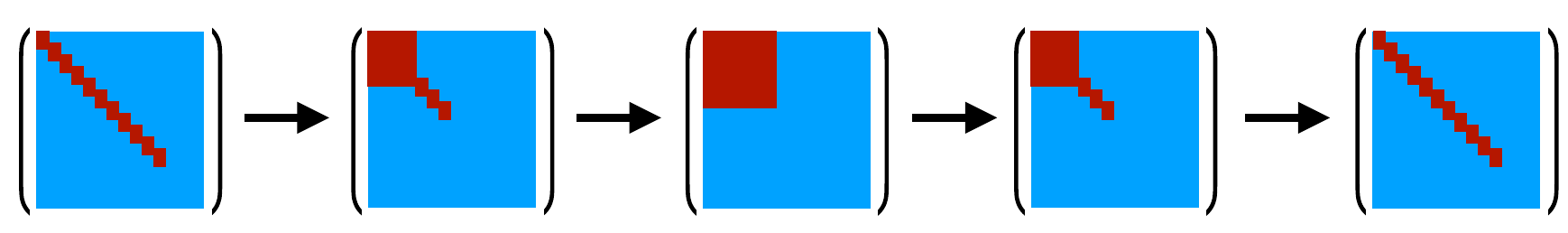}}
\end{center}
\caption{Formation and evaporation of a small black hole, in the second-quantized picture proposed in this article.
Similarly to Matrix Theory~\cite{Banks:1996vh}, the block-diagonal structure is manifest after gauge fixing.  
Small excited blocks describe small objects such as gravitons. 
They merge together and form a large resonance, which is interpreted as a black hole. 
Eventually black hole decays by emitting small objects. 
}\label{fig:life-of-BH}
\end{figure}

What if there are multiple excitations with smaller energy (say $E\sim N^1$ or $N^0$), which are localized in separate locations in the emergent space (and, possibly, also in the spatial dimensions of QFT)? 
Then each of them should clump to a small block in the color space, such that a large enhancement factor appears.
In this way, the block-diagonal structure is favored. 
Note that $Y_I$, rather than $\hat{X}_I$, are block-diagonal. 
It is natural to regard each block as an extended object such as fundamental string or small black hole, that moves in the emergent space and interacts with others.
In the large-$N$ limit, various many-body states can be realized as configurations with multiple excited blocks.
The number of blocks and their sizes can take arbitrary values.
Therefore, the large-$N$ limit of gauge theory can serve as a second quantization of the gravity side. 
In Fig.~\ref{fig:life-of-BH}, a schematic picture of the formation and evaporation of black hole is shown. 
\section{Conclusion and discussion}
The above reasoning is based on generic kinematical features of gauge theory.
To establish a quantitative correspondence between gauge theory and gravity, we have to solve theory-specific strongly-coupled dynamics.
In addition to traditional quantum Monte Carlo simulations (e.g., Refs.~\cite{Berkowitz:2016jlq,Watanabe:2020ufk}),    
new technologies such as deep learning~\cite{Han:2019wue} and quantum simulation~\cite{Gharibyan:2020bab} might be useful. 
If the second-quantization picture is valid indeed, we might be able to decode the hologram and understand nontrivial processes such as the formation and evaporation of black hole. 

\begin{center}
{\bf Acknowledgments}
\end{center}
I would like to thank Hidehiko Shimada and Bo Sundborg for extremely useful comments.  
He was supported by the STFC Ernest Rutherford Grant ST/R003599/1.

\end{document}